**Atomic-scale origin of the low grain-boundary resistance in perovskite solid electrolytes**

*Tom Lee, Ji Qi, Chaitanya A. Gadre, Huaixun Huyan, Shu-Ting Ko, Yunxing Zuo, Chaojie Du, Jie Li, Toshihiro Aoki, Caden John Stippich, Ruqian Wu[*], Jian Luo[*], Shyue Ping Ong[*], and Xiaoqing Pan[*]*

T. Lee, H. Huyan, C. Du, C. J. Stippich, Prof. X. Pan

Department of Materials Science and Engineering, University of California at Irvine, Irvine, CA, USA.

E-mail: xiaoqinp@uci.edu

J. Qi, S. T. Ko, Prof. J. Luo

Materials Science and Engineering Program, University of California San Diego, La Jolla, CA, USA.

E-mail: jluo@alum.mit.edu

C. A. Gadre, J. Li, Prof. R. Wu, Prof. X. Pan

Department of Physics and Astronomy, University of California at Irvine, Irvine, CA, USA.

E-mail: wur@uci.edu

Y. Zuo, Prof. J. Luo, Prof. S. P. Ong

Department of NanoEngineering, University of California San Diego, La Jolla, CA, USA.

E-mail: ongsp@eng.ucsd.edu

Dr. T. Aoki, Prof. X. Pan

Irvine Materials Research Institute, University of California at Irvine, Irvine, CA, USA.

**Abstract**

Oxide solid electrolytes (OSEs) have the potential to achieve improved safety and energy density for lithium-ion batteries, but their high grain-boundary (GB) resistance is a general bottleneck. In the most well studied perovskite OSE, $Li_{3x}La_{2/3-x}TiO_3$ (LLTO), the ionic conductivity of GBs is about three orders of magnitude lower than that of the bulk. In contrast, the related $Li_{0.375}Sr_{0.4375}Ta_{0.75}Zr_{0.25}O_3$ (LSTZ0.75) perovskite exhibits low GB resistance for reasons yet unknown. Here, we used aberration-corrected scanning transmission electron microscopy and spectroscopy, along with an active learning moment tensor potential, to reveal the atomic scale structure and composition of LSTZ0.75 GBs. Vibrational electron energy loss spectroscopy is applied for the first time to characterize the otherwise unmeasurable Li distribution in GBs of LSTZ0.75. We found that Li depletion, which is a major reason for the low GB ionic conductivity of LLTO, is absent for the GBs of LSTZ0.75. Instead, the low GB resistivity of LSTZ0.75 is attributed to the formation of a unique defective cubic perovskite interfacial structure that contained abundant vacancies. Our study provides insights into the atomic scale mechanisms of low GB resistivity and sheds light on possible paths for designing OSEs with high total ionic conductivity.

**1. Introduction**

As the world transitions toward powering itself solely on sustainable sources of energy, energy storage devices become a critical component in making this process a reality. Among these devices, secondary lithium-ion batteries (LIBs) have been regarded as the most dominant and promising energy storage device because of their high energy density, high power density, long cycle life, and low self-discharge. Liquid electrolytes, as the media for ion transportation for current commercial LIBs, still suffers from a few drawbacks, including the safety concerns of their thermal stability and leakage.[1] On the other hand, solid-state electrolytes are intrinsically safer and leak free.

Among the various solid electrolytes, oxide based ones have drawn significant interest because of their electrochemical, thermal, and structural stability. One bottleneck that limits most oxide solid electrolytes from achieving high total ionic conductivity is the large grain-boundary (GB) resistance.[2–10] Many lithium-ion conductors have bulk ionic conductivity that are high enough for building practical all-solid-state battery (ASSB).[3] However, their total ionic conductivity is typically orders of magnitude lower than the bulk ionic conductivity due to large GB resistance.[3] One of the most well studied materials that exhibits this issue is the perovskite type solid electrolyte $Li_{3x}La_{2/3-x}TiO_3$ (LLTO, $0 < x < 0.16$). In its first report by Inaguma and coworkers, LLTO displayed a high bulk ionic conductivity ($\sigma_b$) of $1.0 \times 10^{-3}$ S/cm at room temperature. But because its apparent GB ionic conductivity ($\sigma_{gb}$) was $7.5 \times 10^{-5}$ S/cm, the total ionic conductivity ($\sigma_t$) was $7 \times 10^{-5}$ S/cm at room temperature.[11] Subsequent study revealed that the GBs comprised of severe structural and chemical deviation of about 2–3 unit cells thick, essentially forming a nanoscale $TiO_2$-like insulating interfacial phase.[12,13] Such GBs prohibit the abundance and transport of charge carrier $Li^+$. Nonetheless, few materials are exceptions where the bulk resistance, rather than GB resistance, is the bottleneck that limits the total ionic conductivity.[14–17] One of them, also a perovskite-type solid electrolyte, is $Li_{0.375}Sr_{0.4375}\square_{0.1875}Ta_{0.75}Zr_{0.25}O_3$ (denoted as LSTZ0.75, where 0.75 refers to the value of $y$ in the general formula $Li_{0.5y}Sr_{1-0.75y}\square_{0.25y}Ta_yZr_{1-y}O_3$ and the $\square$ represents the A-site vacancy). It achieved a $\sigma_b$ of $3.5 \times 10^{-4}$ S/cm, a rather high apparent $\sigma_{gb}$ of $1.2 \times 10^{-3}$ S/cm (including the grain size effect), and a $\sigma_t$ of $2.7 \times 10^{-4}$ S/cm at room temperature.[16] However, it is unknown why LSTZ and LLTO show different trends in grain boundary conduction even though they have the same perovskite (ABO$_3$) crystal structure.

Atomic scale characterization of microstructure and grain boundaries is crucial for understanding the unusual GB ionic conductivity enhancement in LSTZ. However, neither the specific GB ionic conductivity $\sigma_{gb}^{spec}$, which represents the intrinsic conduction behavior of the GB, nor the atomic scale microstructures of LSTZ GB and grain bulk have been reported. It is reasonable to expect the specific GB ionic conductivity of LSTZ is higher than those of LLTO and other Li-ion conductors because of the rather high $\sigma_{gb}$. Thus, it is important to study the GB microstructure of LSTZ0.75, as it will unravel the origin of its low GB resistance and provide insights to overcoming the ubiquitous bottleneck of high GB resistance in other materials. It is also significant to study the bulk phase and microstructure of LSTZ0.75, because this will shed light on how to increase the material's bulk, and thus total, ionic conductivity. Additionally, the atomic scale characterization of Li$^+$ ion distribution is important for solid electrolyte materials. Previous studies have shown distinct Li-K edges were difficult to identify in perovskite-type solid electrolytes, owing to their lower volume densities of Li$^+$ ions compared to other types of electrolytes.[18,19] Thus, a technique other than conventional low energy core-loss electron energy loss spectroscopy (EELS) is needed.

Computational studies of complex GB structures have also been constrained by the tradeoff between accurate, but expensive *ab initio* methods[20–25] and cheap, but inaccurate classical interatomic potentials (IAPs).[26] In recent years, IAPs that utilize machine learning (ML) to map the potential energy surface to local environment descriptors have emerged as a promising alternative that combines high accuracy and speed.[26–32] Nevertheless, studies of GB structures with ML-IAPs remain challenging due to the more complex local environment at GB regions,[29,33] with most studies focusing on low-sigma GB models,[34,35] even though high-sigma and general GBs are frequently observed in experiments.

Herein, we present the first atomic scale study of LSTZ0.75 to reveal the origin of its low GB resistivity. Aberration-corrected scanning transmission electron microscopy (STEM) in combination with EELS show that GBs in LSTZ0.75 consist of defective perovskite structures with significant amounts of vacancies. Furthermore, state-of-the-art vibrational EELS technique is employed for the first time to determine the otherwise unmeasurable Li distribution in GBs of LSTZ0.75. Its result shows that $Li^+$ concentration at GBs is the same as that in grain bulk, indicating GBs do not suffer from the detrimental $Li^+$ depletion. Using an efficient active learning strategy, a moment tensor potential (MTP) for LSTZ0.75 was developed to accurately model both bulk and GB structures. Hybrid Monte Carlo/molecular dynamics (MC/MD) simulations corroborate the existence of Sr vacancies in both low- and high-sigma GBs, leading to comparable $Li^+$ diffusivity at the GB and bulk regions. Additionally, these simulations indicate that disruption of A-site ordering in the grain bulk of LSTZ0.75 will result in higher bulk ionic conductivity. Our results provide a new insight of the atomic scale origin of the low GB resistivity in LSTZ0.75, which can help us to design and optimize other solid electrolytes. In a broader perspective, the vibrational EELS approach presented here should be applicable to perovskite oxide solid electrolytes in general as well as any LIB material in which distinct Li-K edges cannot be convincingly identified via conventional low energy core-loss EELS. Similarly, the new active learning workflow we applied here should be generalizable to other complex structures beyond the scale of AIMD simulations.

## 2. Results and Discussion

### 2.1 Structural ordering in LSTZ grain bulk and its effect on bulk ionic conductivity

The LSTZ0.75 ceramics were synthesized via conventional solid state reaction method. Results from crystal structure, ionic conductivity, and electrochemical stability characterization (Figure S1, Supporting Information) demonstrate that our as-synthesized LSTZ0.75 ceramics have bulk properties similar to those reported in literature. A detailed analysis of these results is in the Supporting Information. Fig. 1a displays the atomic-resolution high-angle annular dark-field (HAADF) STEM image of an LSTZ0.75 (010) faceted GB. For the left-side grain in [100] zone axis, there is an alternate stacking between bright and dark A-site atomic columns near the GB that persists approximately 17 unit cells into the grain interior. Fig. 1b and c display the fast Fourier transform (FFT) pattern of the region boxed in blue and red, respectively. Only primary spots corresponding to planes perpendicular to the [100] zone axis are observed in Fig. 1b. However, besides the primary spots, one set of superlattice spots arose from the A-site alternate stacking is also prominent in Fig. 1c. This set of superlattice spots is associated with the ordering in the (010) plane. The representative $(0\ \pm\frac{1}{2}\ 1)$, $(0\ \pm\frac{1}{2}\ 0)$, and $(0\ \pm\frac{1}{2}\ -1)$ spots were circled in green. From the analysis of FFT patterns, it appears that only ordering with long coherence length is present in the region near GB. However, FFT patterns lose phase information and may overlook certain features. Low magnification HAADF-STEM image (Figure S2, Supporting Information) shows numerous dark spots in the grain interior of both grains which can be either a local clustering of point defects or structural ordering with extremely short coherence length. The former would not exhibit superlattice spots in electron diffraction, while the latter would. In order to thoroughly understand this feature, selected area electron diffraction (SAED) in TEM is conducted. Fig. 1d shows SAED pattern of LSTZ0.75 grain bulk along the [100] zone axis. Surprisingly, the SAED collected from grain bulk exhibits two sets of diffraction spots that had failed to be observed in the corresponding FFT pattern. Twelve representative spots from these two sets of diffraction spots were circled in red. Fig. 1e shows SAED pattern of superlattice structure near (010) faceted GB along the [100] zone axis. The SAED from A-site alternate stacking similarly exhibits an additional set of diffraction spots that was not revealed in the corresponding FFT pattern. Six representative spots from the additional set of diffraction spots were circled in red.

Results from SAED reveal the existence of short coherence length ordering in LSTZ0.75 grain bulk. As for the bulk region near (010) faceted GB, both short and long coherence length ordering were present, though the short coherence length ordering in the (001) plane is masked by the long coherence length ordering in the (010) plane. The short coherence length ordering observed here is similar to the mesoscopic coherence length ordering observed in cubic LLTO.[36] Both features could not be detected by XRD due to the short length scale. However, the mesoscopic ordering in cubic LLTO can be detected by FFT pattern (and no SAED was performed to confirm), whereas the short coherence length ordering in cubic LSTZ was overlooked by FFT pattern and can only be detected using SAED.

The structural ordering and local composition were further elucidated using a moment tensor potential (MTP) that has been fitted to accurately reproduce the DFT potential energy surface of both bulk and GB structures of LSTZ0.75 using an active learning scheme (Figure S3, Supporting Information and Methods). Hybrid Monte Carlo/molecular dynamics (MC/MD) simulations were performed on a $2 \times 2 \times 2$ supercell at four different temperatures (298, 723, 1148 and 1573 K). Fig. 2a shows evolution of the LSTZ0.75 unit cell as temperature increases. Between 0 and 723 K, LSTZ0.75 is characterized by A-site alternate stacking of Sr-rich and Sr-poor layers (Fig. 2a). The A-site ordering parameter $S$ is near its maximum value of 0.78 (Fig.

2b), i.e., nearly all Sr are in the Sr-rich layers. However, this A-site stacking becomes disordered at above 1148 K, and $S$ decreases to below 0.2.

MD simulations were performed to the equilibrated structures (Fig. 2a) from the 5 ns MC/MD simulations at the four different temperatures. As shown in the Arrhenius plot in Fig. 2c, simulated bulk ionic conductivity at 300 K ($\sigma_{300K}$) of the two structures equilibrated at or below 723 K with $S$ ~0.78 match perfectly with experimentally measured $\sigma_b$ at room temperature, while the $\sigma_{300K}$ of the other two structures equilibrated at or above 1143 K with disordered A-sites are promoted by 1-2 times. It was also found that a decrease in $S$ leads to larger $\sigma_{300K}$ and lower activation energy ($E_a$). The promoted $\sigma_{300K}$ and the lower $E_a$ match better with the highest experimental $\sigma_b$ of $4.1 \times 10^{-4}$ S/cm at 25 °C and the low $E_a$ of 0.33 eV from 25 to 140 °C reported for a hot-pressed LSTZ0.75 sample, whose A-site alternate stacking was not yet characterized.[37] Similar trends have been observed for the related LLTO material, where higher calcination temperatures lead to a tetragonal to cubic transformation and an increase in $\sigma_b$ from ~$7 \times 10^{-4}$ S/cm to ~$1.5 \times 10^{-3}$ S/cm.[38,39] With the SAED results indicating the existence of short coherence length A-site ordering in LSTZ0.75 grain bulk and the trend uncovered in Fig. 2c, we believe the $\sigma_b$ of LSTZ0.75 can be further enhanced by promoting disorder in A-sites. Meanwhile, the promoted $\sigma_b$ and the decreased $E_a$ of the hot-pressed LSTZ0.75 might be attributed to the elimination of A-site alternate stacking, for which we expect experimental characterization to test out in future.

**2.2 Microstructure of LSTZ0.75 GBs revealed by STEM-EELS**

Next, we investigate the GBs of LSTZ in order to reveal the origin of its low GB resistance. Electron Backscatter Diffraction (EBSD) results (Figure S4, Supporting Information) indicated that the average grain size is 3.38 ± 1.13 μm, and the GBs mainly consist of randomly orientated grains. In total, we have imaged more than seven hundred GBs, of which ~12% have (010) faceted grain terminal surfaces on one side. As shown in Fig. 1a, an obvious dark band can be observed at the (010) faceted GB. Fig. 1f and 1g exhibits atomic-resolution HAADF-STEM images of general GBs with complex, high-index surfaces. Obvious dark bands at the GBs are still present, while no obvious ordering is observed in the region near GBs. In fact, of the >700 GBs imaged, we have only observed the alternate stacking between bright and dark A-site atomic columns near (010) faceted GBs (~12% of all GBs). However, all of the GBs exhibited a drastic decrease in image intensity in comparison with grain bulk, regardless of the relative orientation between the two adjacent grains. Since the HAADF-STEM image intensity is dictated by the average atomic number of the elements present,[40] this observation clearly indicated a compositional variation across the GB. Specifically, the decrease in intensity at the A-site atomic columns can be attributed to a decrease in concentration of Sr (either an increase in the amount of Li substituting Sr or simply increase in vacancies). Similarly, the decrease in intensity at the B-site atomic columns can be attributed a decrease in concentration of Ta and/or Zr. It is worth noting that despite the compositional change, the cubic perovskite crystal structure is still maintained at the GBs. Figure S5a and S5b of the Supporting Information show high magnification atomic-resolution HAADF-STEM images of the general GB from Fig. 1g and a (010) faceted GB (with respect to the right-side grain), respectively. Since the same crystal lattices from the bulk extend all the way to the GBs, and the only change is the observed decrease in intensity at GBs, crystal structure of the GBs is also cubic perovskite.

To better understand the observed features in HAADF images, high spatial resolution STEM-EELS measurements were performed at (010) faceted GBs and general GBs. Fig. 3a and 3b show the integrated EEL spectra of O-K edge and edges in the high energy loss regime (Sr-

$L_{2,3}$, Zr-$L_{2,3}$, Ta-$M_{2,3}$, and Ta-$M_{4,5}$ edges), respectively. Atomic-resolution HAADF-STEM image of the (010) faceted GB from which the EEL spectra were collected is shown in Fig. 3c. The GB again exhibited a dramatically lower image intensity in comparison with the bulk. For the left-side grain, the alternate stacking between bright and dark A-site atomic columns, which extends ten unit cells from GB into grain interior, is also observed. Fig. 3d, e, f, and g display the atomic-resolution elemental maps of Sr, Ta, Zr, and O, respectively. The maps were generated using intensities from Sr-$L_{2,3}$, Zr-$L_{2,3}$, Ta-$M_{4,5}$, and O-K edges. As expected, the elemental maps (Fig. 3 d, e, and f) show Sr atoms reside in the A-site positions, and Ta/ Zr atoms reside in the B-site positions. Atoms of a single unit cell are labeled, in accordance with the schematic of LSTZ0.75 crystal structure (Figure S1a, Supporting Information), in Fig. 3c. Bright field (BF) STEM image (Figure S6b, Supporting Information) further confirms that, in addition to overlapping with B-site atoms, O atoms also reside between two adjacent B-site atoms. In summary, the STEM images and EELS elemental maps confirmed the proposed schematic of LSTZ0.75 crystal structure (Figure S1a, Supporting Information) to be correct.

To analyze the correlation between HAADF image and the elemental maps, a plot of vertically integrated intensity profiles across the GB is shown above the HAADF image (Fig. 3c). For the GB, the dramatic decrease in image intensity correlates well with the decrease in intensity of Sr, Ta, and Zr signals, though the decrease in Zr signal is much less than the other two. Specifically, there is a ~26%, 11%, and 7% decrease in Sr, Ta, and Zr signals, respectively, at the GB when compared to those from the abutting grains. This indicates that the GB contains more Sr, Ta, and Zr vacancies than the grain bulk, and its crystal structure resembles that of a defective cubic perovskite. Further analysis reveals that higher Sr and Ta signals are observed at the left side adjacent to the GB. For the left-side grain, variation in HAADF intensity correlates reasonably well with that in Sr elemental map intensity. The bright A-site atomic columns in superlattice (SL) region contains more Sr than those in other regions of the grain and can be considered as Sr-rich layer. By the same token, the first dark A-site atomic column immediately adjacent to the GB core contains less Sr than A-site atomic columns in other regions of the grain and can be considered as a single Sr-poor layer. The remaining dark A-site atomic columns contain approximately the same amount of Sr as those in other regions of the grain. Analysis of the Sr signals indicates that this SL region contains more Sr than the rest of grain bulk. In other words, this region has a higher density and the A-site vacancies are filled with Sr. This decrease of A-site vacancies will decrease the number of Li percolation pathways and impede $Li^+$ ion migration. Therefore, from both structural ordering and concentration of A-site vacancy point of view, the superlattice structure is unfavorable for achieving high bulk ionic conductivity. Nonetheless, since the (010) faceted GBs only constitute ~12% of all GBs, the impact such structure has on the overall bulk ionic conductivity should be insignificant. Finally, slightly higher Ta signals can be observed in the three B-site atomic columns located at -0.5, -0.9, and -1.3 nm distance to GB. The results above indicate that the GB has undergone elemental segregation, with Sr and Ta segregates out of the GB core and into side of the left grain. The increased amount of cation vacancies at GB provides more Li percolation pathways, which will facilitate $Li^+$ ion migration and decrease GB resistance. Unlike the intensity of Sr, Ta, and Zr signals, no decrease is observed in the intensity of O signals at the GB. Finally, STEM imaging coupled with EELS measurements was also performed at general GBs and results are shown in Figure S7 of the Supporting Information. A similar trend is observed in the atomic-resolution HAADF image and elemental maps of Sr, Ta, Zr, and O. The general GB contains Sr, Ta, and Zr vacancies. But instead of Zr vacancies, Ta vacancies is much less than the other two at this GB.

Obvious elemental segregation of Sr and Zr can be observed at both sides adjacent to the GB. The differences between (010) faceted GB and general can be attributed to subtle variation in the composition of GBs. Consistent with the lack of ordering observed in the HAADF-STEM images (Fig. 1f and 1g), the Sr elemental map did not exhibit alternate stacking of A-site.

**2.3 LSTZ0.75 GB composition and its effect on ionic conductivity revealed by MTP**

The relationship between GB composition and ionic conductivity was probed using the MTP. Specifically, four low-sigma GBs (symmetric tilt $\Sigma5[100](0\bar{1}2)$, simple twist $\Sigma5[100](100)$, symmetric tilt $\Sigma3[110](1\bar{1}1)$ and simple twist $\Sigma3[110](110)$) and one high-sigma (tilt $\Sigma51[110](001)(3\bar{3}10)$) GBs were used as model systems (Fig. 4a). The MTP successfully reproduced the GB energies of the low-sigma GBs $\gamma_{GB}$ to within 0.10 J/m$^2$ (Table S1, Supporting Information). In general, the GBs of LSZT0.75 are predicted to have significantly higher energies (0.67-0.74 J/m$^2$) than LLTO (0.30 J/m$^2$),[41] which may result in smaller GB regions and partially account for lower GB resistivity. MC/MD simulations were performed at the experimental calcination temperature of 1573 K. It is found that the atomic percentage of Li$^+$ at GB regions increased from around 8% to over 11% in the two simple twist GBs and the high-sigma GB, i.e., $\Sigma5[100](100)$, $\Sigma3[110](110)$ and $\Sigma51[110](001)(3\bar{3}10)$ (Fig. 4b and Table S2, Supporting Information). This is accompanied by a decrease in the atomic percentage of Sr, with the formation of more Sr vacancies, at the GB regions. From these results, Li$^+$ depletion, one of the major causes of low GB conductivity in LLTO,[12] is generally unfavorable in LSTZ, while formation of Sr vacancies at GB regions is favorable, in line with the experimental observations of Sr vacancies at GB regions. MD simulations were conducted to the two most stable low-sigma GBs (symmetric tilt $\Sigma5[100](0\bar{1}2)$, simple twist $\Sigma3[110](110)$) and the high-sigma tilt $\Sigma51[110](001)(3\bar{3}10)$ GB. As shown in Fig. 4c, non-Arrhenius behaviors with transition temperatures of around 600 K can be generally observed, similar to what has been observed for LLTO.[11,42] However, the GB Li$^+$ diffusivities $D_{Li, 300K}$ of LSTZ0.75 are only 2-3 times lower than those of bulk LSTZ0.75 (Table S3, Supporting Information). This is in sharp contrast to LLTO, where it has been reported that the GB ionic conductivity is 1-2 orders of magnitude lower than bulk ionic conductivity.[11,41] The GB activation energies in LSTZ0.75 are also comparable to the bulk value, while in LLTO, the GB activation energies are generally higher than bulk.[41] Furthermore, we note that the non-equilibrated GBs have a higher $E_a$ and lower $D_{Li, 300K}$ (Table S3, Figure S8, Supporting Information), which indicates that Li enrichment and Sr vacancies play a role in increasing the GB Li$^+$ diffusivity. Most importantly, these trends were consistently observed in the low-sigma GBs and the high-sigma GB, indicating fast Li$^+$ diffusion at low-sigma GBs as well as general GBs of LSTZ0.75. We note that these general observations remain valid with and without equilibration.

**2.4 Li distribution in LSTZ0.75 microstructure revealed by vibration EELS**

Although attempts were made to investigate the Li distribution at the GBs by identifying the Li-K edge, the low Li content in LSTZ0.75, poor scattering power of Li ($Z = 3$), and overlap with the Ta-O$_{2,3}$ edge makes this investigation challenging. More detailed results are reported in Figure S9 of the Supporting Information. Representative spectra in Fig. 5a hardly shows distinguishable features for Li-K edge and is overwhelmed by the Ta-O$_{2,3}$ edges due to larger quantities of Ta. Thus, the variation of the average peak height indicated in Fig. 5b includes a combination of Ta-O$_{2,3}$ and Li-K edge signals. The line profile and contour plots (Figs. 5b, d, respectively) show two small humps in the left grain owing to the SL and a dip at the GB which may suggest Ta or Li vacancies but ambiguous as to which is most responsible for the decrease in signal. As a result, mapping of the Li distribution in grains and GBs of LSTZ0.75 via the Li-K edge is not feasible in systems which contain a large Ta:Li ratio.

To resolve this issue, we map the Li distributions by probing Li-O vibrations. To this end, we employ a state-of-the-art vibrational EELS technique enabled by high energy resolution monochromation which has facilitated the spectroscopy of vibrational excitations at the nanometer[43] and even atomic scales.[44,45] We employ a dark field vibrational EELS (DF VibEELS) beam-detector geometry (Figure S10, Supporting Information) to ensure only high spatial resolution signals are acquired.[44,46]

Representative background-subtracted spectra of LSTZ in the grain interior, grain boundary, and SL region spanning 10-150 meV (Fig. 5e) is consistent with the Raman spectrum of the LSTZ pellet (Figure S11c, Supporting Information). Heavier elements in the A- and B-sites, namely Sr, Ta, and Zr, contribute to lower energy vibrations with oxygen and comprise of the low energy peaks from 10 to 50 meV while the lighter Li vibrates with O at a higher frequency and produces vibrations in the 50 to 100 meV range (Figure S11b, Supporting Information). Several striking features can be observed in the contour plot (Fig. 5h), including a strong variation in intensity in the SL region and a drastic decrease in intensity at the GB in the low energy region of the phonon signal. The orange curve in Fig. 5f, containing mainly Sr, Ta vibrations with O (Figure S11b, Supporting Information), shows strong intensity modulation in the SL region due to the increased concentration of Sr in the SL region while the dip at the GB suggests deficiencies in Sr, Ta vibrational species. Given that the concentration of O is constant throughout this region (Fig. 3g), the DF VibEELS mapping (Fig. 5f) shows Sr and Ta vacancies and is consistent with Figs. 3d-f. The green curve represents the spatial trend of vibrations that are largely dominated by vibrations consisting of Ta, Zr. Here the dip at the GB is less drastic owing to slight vacancies in Ta and Zr which can be more clearly seen in Fig. 5h. A similar contrast for Ta, Zr-O vibrations in the SL region is seen in Fig. 5f and, to a much lesser degree, in the intensity profiles in Fig. 3, which suggests that DF VibEELS is extremely sensitive to local elemental modulation and can be extended to map minute variations in composition. This is not unlike VibEELS ability to measure structural changes that manifest as defect modes.[47] The purple curve, containing a majority of Li-O vibrations, also shows strong intensity modulation in the SL region produced by an alternating Sr-rich and Li-rich SL. Unlike the other curves, Li-O shows no decrease in integrated intensity at the GB, indicating that there are no Li vacancies at the GB. This can also be seen in the representative spectra in Fig. 5e where the intensities in the 50-100 meV region hardly change between the bulk and the GB. Given there are Sr, Ta, and Zr vacancies, the relative atomic % of Li is higher at the GB, which is consistent with Fig. 4b and Table S2 of the Supporting Information. Coupled with the phonon DOS, the DF VibEELS mapping allows us to not only map the distribution of Li but also Sr, Ta, and Zr at the atomic scale. This technique is indispensable for solid electrolytes with compositions that contain large Ta:Li ratios, as the Ta-$O_{2,3}$ edge completely obscures and overwhelms the Li-K edge and makes it impossible to map Li distributions by conventional core-loss EELS. As a result, DF VibEELS technique emerges as a powerful technique that is solely capable of characterizing LSTZ and LSTZ derivatives (*e.g.*, $Li_{0.375}Sr_{0.4375}Hf_{0.25}Ta_{0.75}O_3$ and $Li_{0.38}Sr_{0.44}Ta_{0.7}Hf_{0.3}O_{2.95}F_{0.05}$) at the atomic scale.

With $Li^+$ ion charge carriers preserved, cubic perovskite structure maintained, and vacancies added at the GBs, $Li^+$ ion transport at the GBs is enhanced in LSTZ0.75. In comparison, $Li^+$ ions were depleted, crystal structure was dramatically altered, and no vacancies were created at the GBs for LLTO[12]. As a result, Li$^+$ ion transport at the GBs is severely impeded and LLTO has a large GB resistance.

## 3. Conclusion

Based on the results presented above, vacancy engineering of the GBs should be a viable approach to increase the GB ionic conductivity of perovskite oxide solid electrolytes. Previous studies have established that ionic conduction in LLTO occurs via the migration of Li$^+$ ions to nearby A-site vacancies.[48–50] Our simulation results indicate the ionic conduction mechanism of LSTZ0.75 is similar to that of LLTO. Since LSTZ0.75 maintained the perovskite framework at GBs, the additional Sr vacancies at its GBs naturally increases the number of percolation pathways and facilitates Li-ion transport. Meanwhile, preservation of the structural framework is important, for it prevents the GBs from forming structures that generate new barriers for Li-ion transport. One method to realize such goals might be incorporation of appropriate dopants into crystal lattice. It has been reported that appropriate crystal doping can improve overall ionic conductivity by introducing additional vacancies while preserving the desired crystal structure.[15,51–55] Beyond this particular LSTZ system, our proposed approach should also benefit other Li-ion conductors, as diffusion to nearby structural defects is a common Li-ion conduction mechanism for solid electrolytes.[55,56]

In conclusion, we report the first atomic-scale study of LSTZ to reveal its microstructures and GB structures to understand their effects on Li-ion conductivity. MTP enabled MC/MD simulations on sophisticated LSTZ0.75 supercells show that disruption of A-site ordering in grain bulk results in increased bulk ionic conductivity. Atomic-resolution STEM and EELS analysis was employed to further study the GB structures. Our results demonstrate that even though the GBs contain substantial amounts of A- and B-site vacancies, the perovskite framework is maintained. DF VibEELS was employed to map the atomic scale $Li^+$ distribution in LSTZ0.75 for the first time, which showed that $Li^+$ concentration at GBs is the same as that inside the bulk phase. MC/MD also reveals a high relative amount of Sr vacancies and $Li^+$ ions at the GBs; these Sr vacancies are found to promote Li-ion diffusivity at GBs. Based on these results, we suggest vacancy and defect engineering as an effective approach to improve GB ionic conductivity of solid Li-ion conductors, given that the material's original structural framework should be maintained. Our study showcases the importance of understanding bulk structural ordering and GB structures in order to improve the macroscopic properties of polycrystalline materials. In addition, we have demonstrated the necessity of using novel, cutting-edge characterization and modeling methods to uncover GB structures and kinetics that would otherwise be impossible to probe with conventional tools, which can be generally applied to other materials.

**Acknowledgements**
T. Lee, J. Qi, and C.A. Gadre contributed equally to this work. This work was funded by the UC Irvine MRSEC, Center for Complex and Active Materials, under National Science Foundation award DMR-2011967. All the TEM experiments were conducted using the facilities at the Irvine Materials Research Institute (IMRI) in University of California, Irvine. The EBSD experiments were performed at UC San Diego Nanoengineering Materials Research Center (NE-MRC). The computing resources are provided by the National Energy Research Scientific Computing Center (NERSC), a U.S. Department of Energy Office of Science User Facility at Lawrence Berkeley National Laboratory and the Extreme Science and Engineering Discovery Environment (XSEDE), which is supported by National Science Foundation grant number ACI-1548562.

**Author contributions**
X.P., S.P.O. and J.L. conceived this project and T.L., J.Q. and C.G. designed the studies. T.L. performed STEM and core-loss EELS experiments and data analysis with the help of C.G. and

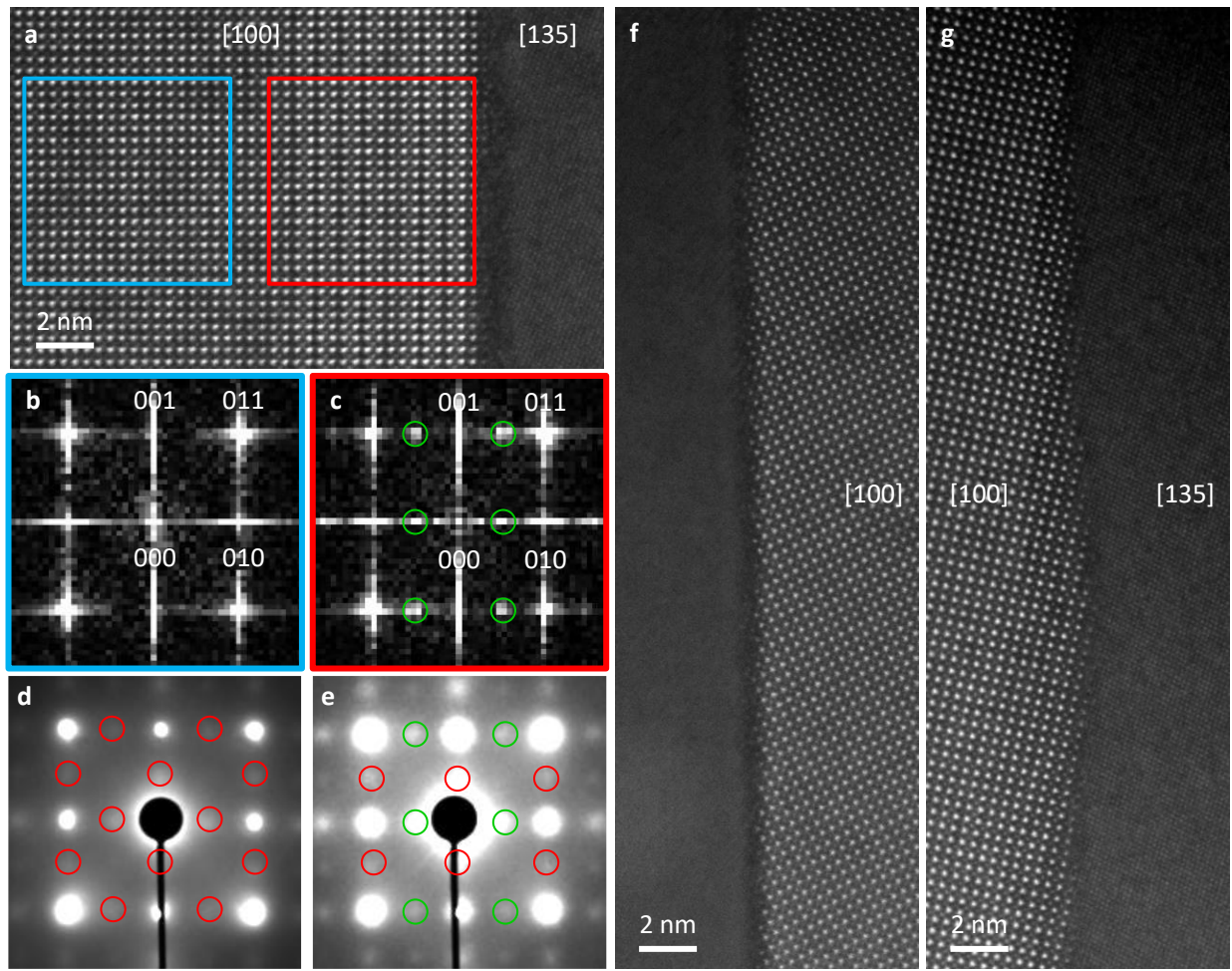

**Figure 1. Atomic-scale study of the crystal structure inside the grain bulk, a (010) faceted grain boundary (GB), and a non-faceted general GB.** a) Atomic-resolution HAADF-STEM image of a (010) faceted GB (with respect to the left-side grain). The zone axes parallel with the incident electron beam were indicated in **a**, **f**, and **g**. b) FFT pattern of the crystal structure inside the grain, boxed in blue in **a**, and c) FFT pattern of superlattice structure near (010) faceted GB, boxed in red in **a**. The $(0\ \pm\frac{1}{2}\ 1)$, $(0\ \pm\frac{1}{2}\ 0)$, and $(0\ \pm\frac{1}{2}\ -1)$ superlattice spots associated with the long coherence ordering in the (010) plane were circled in green. d) SAED pattern of LSTZ0.75 grain bulk along the [100] zone axis. The $(0\ \pm\frac{1}{2}\ 1)$, $(0\ \pm\frac{1}{2}\ 0)$, $(0\ \pm\frac{1}{2}\ -1)$, $(0-1\ \pm\frac{1}{2})$, $(0\ 0\ \pm\frac{1}{2})$, and $(0\ 1\ \pm\frac{1}{2})$ diffraction spots failed to be observed in the corresponding FFT pattern were circled in red. e) SAED pattern of superlattice structure near (010) faceted GB along the [100] zone axis. The $(0-1\ \pm\frac{1}{2})$, $(0\ 0\ \pm\frac{1}{2})$, and $(0\ 1\ \pm\frac{1}{2})$ diffraction spots failed to be observed in the corresponding FFT pattern were circled in red. f), g) Atomic-resolution HAADF-STEM images of general GBs. The left-side grain in **f** was not oriented along any particular zone axis.

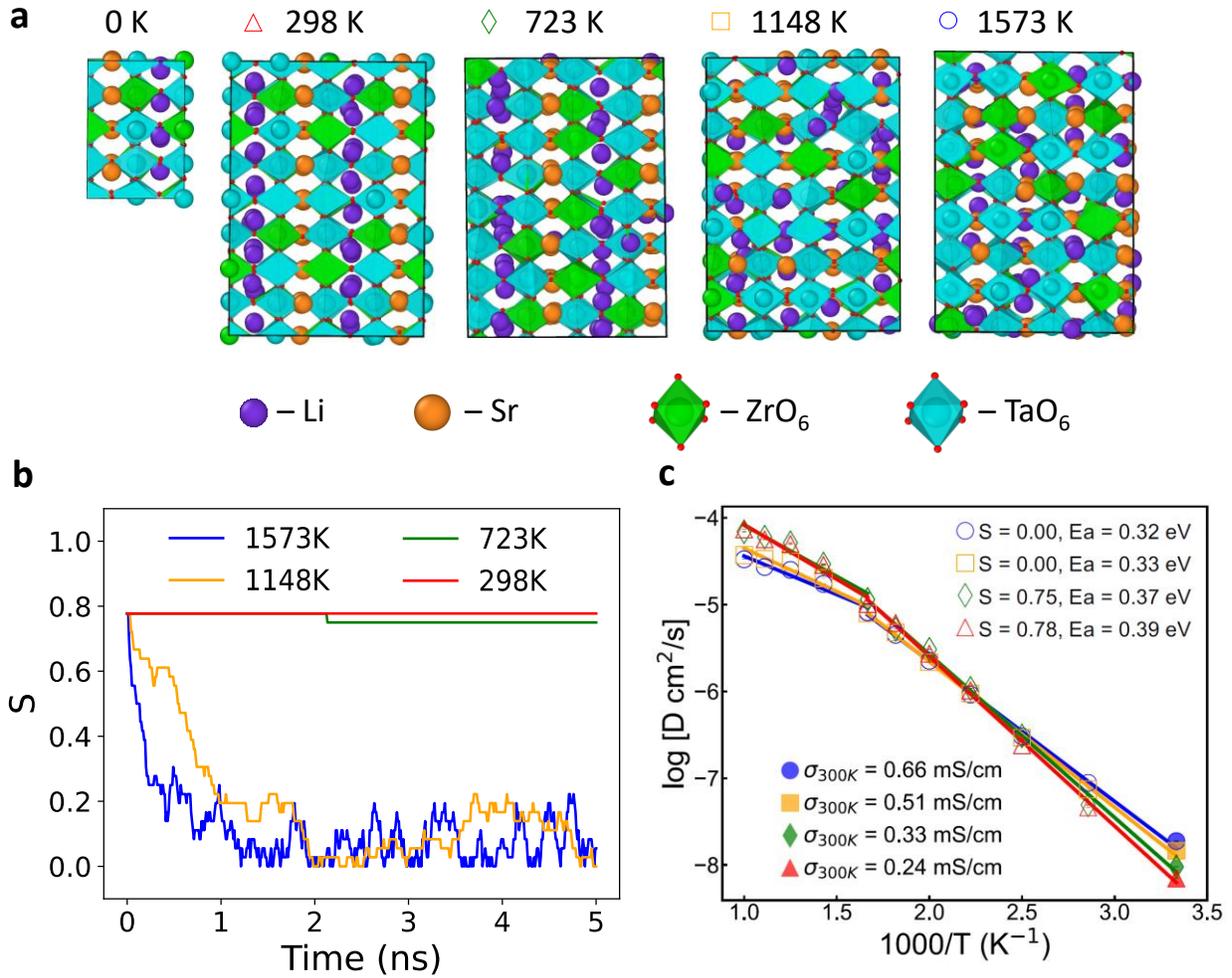

**Figure 2. The temperature-dependent A-site ordering in bulk LSTZ0.75 and its effect on Li⁺ diffusion.** a) Crystal structures of LSTZ0.75 unit cell with the most stable ordering by DFT at 0 K and the 2 × 2 × 2 supercells equilibrated by MC/MD with MTP at 298 to 1573 K with 425 K intervals. b) The evolution of A-site ordering parameter $S$ during the 5 ns MC/MD simulations at the four temperatures. The A-site ordering parameter is given by $S = \frac{R_{Sr-rich} - R_{Sr-overall}}{1 - R_{Sr-overall}}$, where $R_{Sr-rich}$ and $R_{Sr-overall}$ refer to A-site occupancy by Sr in Sr-rich layers and the overall A-site occupancy by Sr in LSTZ. For LSTZ0.75, $S$ ranges from 0 (complete disorder) to 0.78 (all Sr in Sr-rich layers). c) The Arrhenius plot of the four models equilibrated at different temperatures. The room temperature ionic conductivity ($\sigma_{300K}$), $S$ and activation energy ($E_a$) below 600 K were listed.

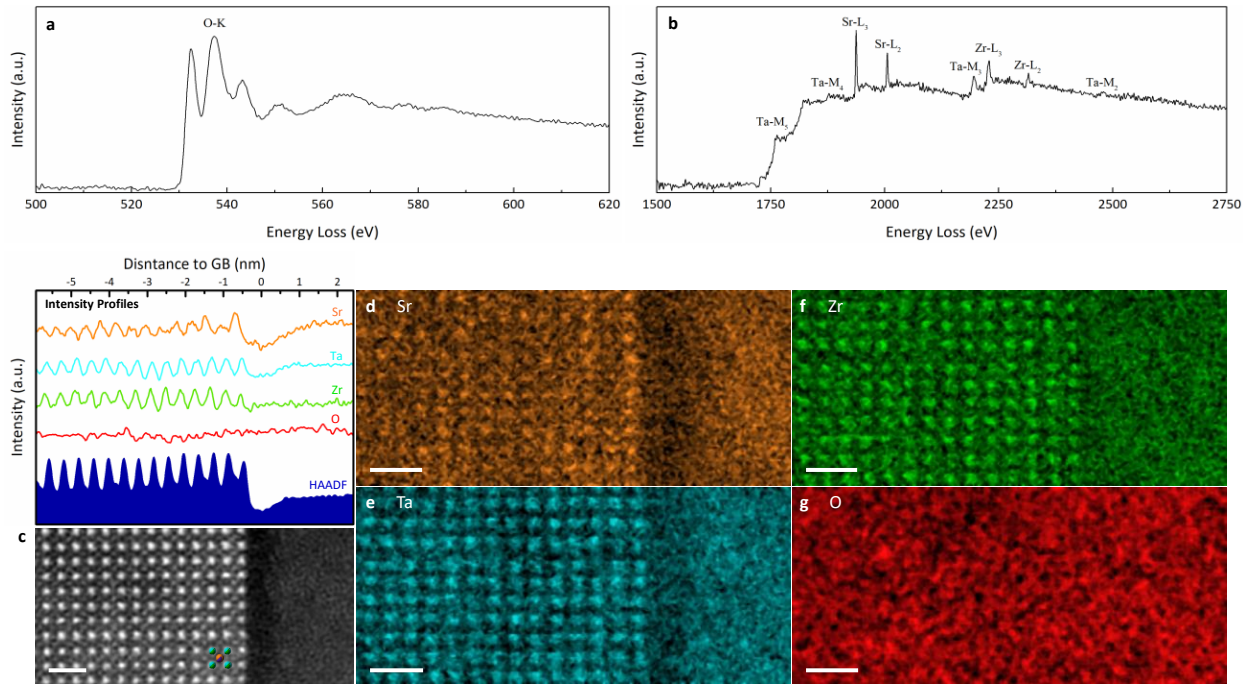

**Figure 3. Core-loss EELS data of (010) faceted GB.** Integrated EEL spectra of a) O-K and b) Sr-$L_{2,3}$, Zr-$L_{2,3}$, Ta-$M_{2,3}$, and Ta-$M_{4,5}$ edges for 010 faceted grain boundary shown in **c**. c) Atomic-resolution HAADF-STEM image of a 010 faceted grain boundary (with respect to the right-side grain). Atoms of a single unit cell (A-site centered view) are labeled, in accordance with the schematic of LSTZ crystal structure (Fig. 1b). Elemental maps of d) Sr, e) Ta, f) Zr, and g) O. All scale bars are 1 nm. Intensity profiles of **c** – **g** are shown above **c**.

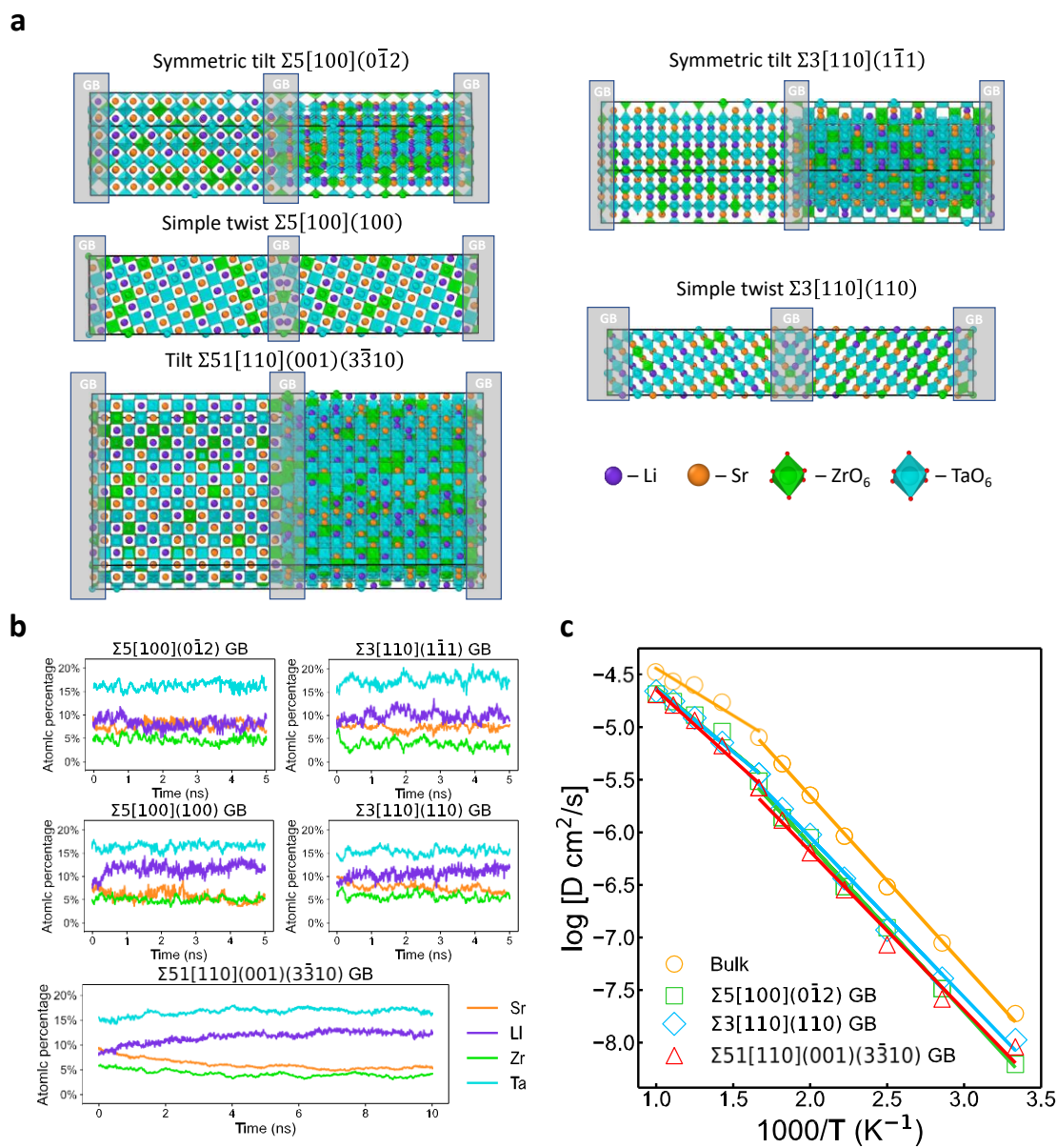

**Figure 4. Local composition and Li+ diffusion at low-sigma and high-sigma GBs.** a) The structures of the four low-sigma and the one high-sigma GB models, where the GB orientations and the GB regions located at the center and edge of the cells were labeled. GB thickness was defined as two or four times of the interplanar distance between GB planes with a general requirement of $d_{GB} > 5$ Å. The exact geometric information of those GB models was provided in Table S1 of the Supporting Information. b) The evolution of the atomic percentage of cations at GB regions of the five GB models during the MC/MD at 1573 K. c) The Arrhenius plot of Li+ diffusivity calculated using the bulk model, the two most stable low-sigma GBs and the high-sigma GB. The GB models were equilibrated with MC/MD simulations at 1573 K. The respective $D_{Li, 300K}$ and $E_a$ were provided in Table 1. The Arrhenius plot of GB models without equilibration was provided as Figure S8 of the Supporting Information.

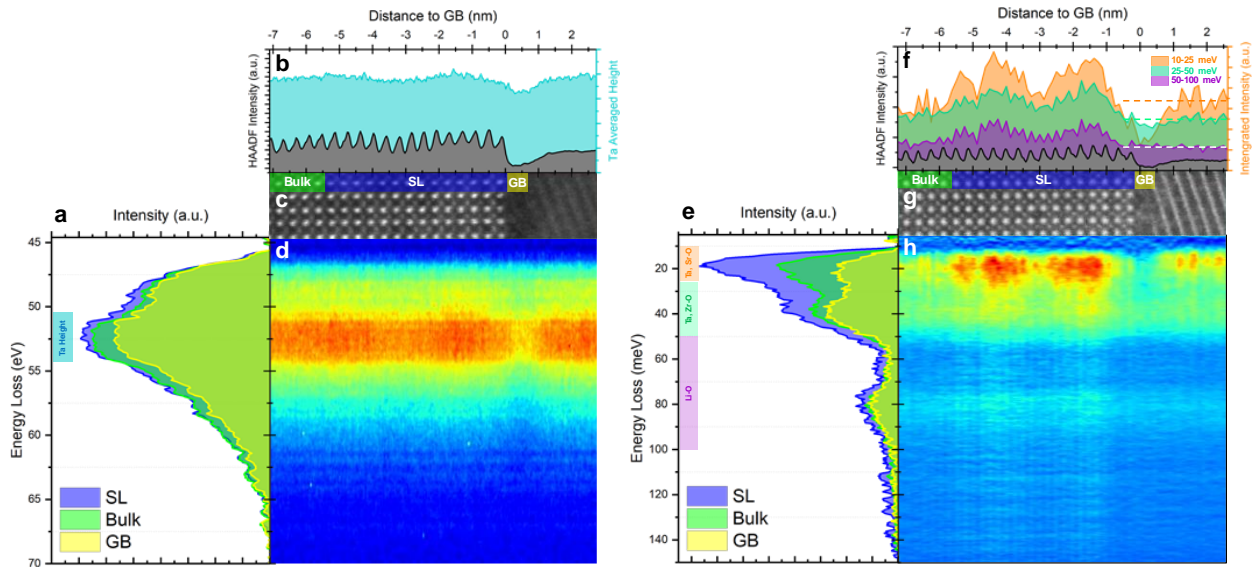

**Figure 5. Low-loss and vibrational EELS of (100)-faceted GB.** a) Representative spectra of Ta-$O_{2,3}$ edge at the SL, bulk, and GB regions. Average Ta-$O_{2,3}$ height is higher at the SL region than both the bulk and GB. The cyan-colored label above the peak denotes the energy range over which the Ta-$O_{2,3}$ height was averaged. b) Line profile of the averaged Ta-$O_{2,3}$ height overlaid with vertically integrated HAADF line profile. C) HAADF image of the region corresponding to the region low energy core-loss EELS was acquired. d) Contour plot of spectral slices stacked horizontally. The vertical axis denotes energy loss while the horizontal axis denotes the horizontal real-space position corresponding to the HAADF in **c**. e) Representative vibrational spectra at the SL, bulk, and GB regions. Averaged vibrational peak is higher at the SL region than both the bulk and GB. The colored labeling above the peak denotes the energy ranges over which the signal was averaged and corresponds to specific elemental vibrations with oxygen. f) Line profiles of the averaged intensities denoted in **e** overlaid with vertically integrated HAADF line profile. While all intensity-averaged curves show modulations in the SL region, only the Ta, Sr-O vibrations show a large dip at the GB. The B-site vibration (Ta, Zr-O) shows a much smaller dip while the vibrational EELS signal corresponding to Li-O vibrations shows no change from bulk at the GB. g) HAADF image of the region corresponding to the region low energy core-loss EELS was acquired. h) Contour plot of spectral slices stacked horizontally. The vertical axis denotes energy loss while the horizontal axis denotes the horizontal real-space position corresponding to the HAADF in **g**.